\documentclass[preprint,prd,superscriptaddress]{revtex4}
\usepackage{amsmath,amssymb}
\usepackage{graphicx}
\usepackage{dcolumn}
\usepackage{bm}
\newcommand{\p}{\partial}
\newcommand{\pslash}{p\kern-1ex /}
\newcommand{\lslash}{l\kern-1ex /}
\newcommand{\kslash}{k\kern-1ex /}
\newcommand{\dslash}{\p\kern-1.2ex /}
\newcommand{\Dslash}{{\cal D}\kern-1.5ex /}
\newcommand{\Aslash}{A\kern-1.2ex /}

\newcommand{\tr}{{\rm tr}}

\newcommand{\Dodwf}{\mathcal{D}}
\newcommand{\bea}{\begin{eqnarray}}
\newcommand{\eea}{\end{eqnarray}}

\newcommand{\BAN}{\begin{eqnarray*}}
\newcommand{\EAN}{\end{eqnarray*}}
\def\u{{\bf u}}
\def\d{{\bf d}}
\def\s{{\bf s}}
\def\c{{\bf c}}
\def\b{{\bf b}}
\def\t{{\bf t}}
\def\q{{\bf q}}

\def\cbar{\bar{\bf c}}

\def\Qbar{\bar{\bf Q}}
\def\Id{ \mbox{1\hspace{-1.2mm}I} }
\begin{document}

\newcommand{\NTU}{
  Physics Department, National Taiwan University, Taipei~10617, Taiwan  
}

\newcommand{\CQSE}{
  Center for Quantum Science and Engineering, National Taiwan University, Taipei~10617, Taiwan  
}

\newcommand{\NTNU}{
  Physics Department, National Taiwan Normal University, Taipei~11677, Taiwan  
}

\preprint{NTUTH-17-505A}
 
\title{Lattice QCD with $N_f = 2+1+1 $ domain-wall quarks}

\author{Yu-Chih~Chen}
\affiliation{\NTU}

\author{Ting-Wai~Chiu}
\affiliation{\NTU}
\affiliation{\CQSE}
\affiliation{\NTNU}

\collaboration{TWQCD Collaboration}
\noaffiliation

\pacs{11.15.Ha,11.30.Rd,12.38.Gc}

\begin{abstract}

We perform hybrid Monte Carlo simulation of (2+1+1)-flavors lattice QCD with the optimal domain-wall fermion  
(which has the effective 4D Dirac operator exactly equal to the Zolotarev optimal rational approximation 
of the overlap Dirac operator).
The gauge ensemble is generated on the $32^3 \times 64 $ lattice with the extent $ N_s = 16 $ 
in the fifth dimension, and with the plaquette gauge action at $ \beta = 6/g^2 = 6.20 $. 
The lattice spacing ($ a \simeq 0.063 $~fm) is determined by the Wilson flow, using the value 
$ \sqrt{t_0} =  0.1416(8) $~fm obtained by the MILC Collaboration for the $(2+1+1)$-flavors QCD.
The masses of $\s$ and $\c$ quarks are fixed by the masses of the vector mesons
$ \phi(1020) $ and $ J/\psi(3097) $ respectively; while 
the mass of the $\u/\d$ quarks is heavier than their physical values, 
with the unitary pion mass $ M_\pi \simeq 280$ MeV (and $ M_\pi L \simeq 3 $). 
We compute the point-to-point quark propagators, and measure the time-correlation functions of 
meson and baryon interpolators.
Our results of the mass spectra of the lowest-lying hadrons containing $ \s $ and $ \c $ quarks 
are in good agreement with the high energy experimental values, together with the predictions 
of the charmed baryons which have not been observed in experiments.   

\end{abstract}

\maketitle

\section{Introduction}

Since the discovery of the Higgs scalar in 2012, the Standard Model (SM) emerged in mid 1970s 
looks to be complete in the sense that all major predictions of the SM have been realized in high
energy experiments, and almost all high energy experimental data 
can be understood in the framework of SM, except the matter-antimatter asymmetry 
and the origin of the neutrino masses. 
Currently, the challenge of high energy physics is to find out whether there is 
any new physics beyond the SM, in view of the generation puzzle and the large number 
of parameters in the SM, which suggest that the SM is an effective theory at the scale 
probed by the present generation of high energy accelerators.  
In order to identify any discrepancies between the high energy experimental 
results and theoretical values derived from the SM, the latter have to be
obtained in a framework which preserves all essential features of the SM. 
Otherwise, it is difficult to determine whether such a discrepancy 
is due to new physics, or just the approximations (or models) one has used. 
So far, the largest uncertainties in the theoretical predictions
of the SM stem from the sector of the strong interaction, namely, QCD. 
Theoretically, lattice QCD is the most viable framework to tackle QCD 
nonperturbatively from the first principles. However, in practice, 
it is difficult to simulate dynamical $ \u$, $\d$, $\s$, $\c$, and $\b$ quarks 
at their physical masses (ranging from $\sim 3-4500 $~MeV), 
in a sufficiently large volume and small enough lattice spacing 
such that the finite-volume and discretization errors are both well under control. 
Note that the $\t$ quark can be neglected in QCD simulations 
since it is extremely short-lived and it decays to W-boson and 
$ \b/\s/\d $ quarks before it can interact with other quarks through the gluons.   
Even after neglecting the $\t$ quark, to simulate $ \u$, $\d$, $\s$, $\c$, and $\b$ quarks 
at their physical masses is still a very challenging problem. 
For example, if one designs the simulation close to the physical pion mass with $ M_\pi \simeq 140 $ MeV 
and $ M_\pi L > 6 $ (to keep finite-volume error under control), 
then it would require a lattice of size $ \sim 100^4 $ to accommodate physical $\c$ quark 
with sufficiently small discretization error, not to mention the much heavier $ \b $ quark. 
The current generation of supercomputers with $ \sim 100 $~Petaflops seems to be marginal for this purpose, 
and the next generation of supercomputers with Exaflops is required to simulate   
$(\u, \d, \s, \c, \b)$ quarks at their physical masses. 

With our rather limited resources, we can only afford to perform lattice QCD simulations 
with domain-wall quarks on a $ 32^3 \times 64 $ lattice, using a GPU cluster with 64 Nvidia GTX-TITAN GPUs. 
Now, even after neglecting the dynamical $ \b $ quark, we still have two options. 
One way is to neglect the $ \c $ quark, and simulate $ (2+1) $-flavors QCD on a coarse lattice 
such that $ M_\pi \sim 140 $~MeV and $ M_\pi L > 3 $, for studying the phenomenology involving the light quarks.  
Instead, we simulate $(\u, \d, \s, \c)$ quarks with sufficiently fine lattice spacing 
satisfying $ m_c a < 1 $, for studying the charm physics, which in turn must render the unitary pion  
[$ m_{u/d} $ (valence) = $ m_{u/d}$ (sea)] heavier than $ 140 $~MeV such that $ M_\pi L > 3 $ 
to avoid large finite-volume error.    

Even with unphysically heavy $\u/\d$ quarks in the sea, the mass spectra of hadrons 
containing $\c$ and $\s$ quarks may turn out to be in good agreement with high energy experimental results, 
as we have observed in our previous studies, for $ N_f = 0 $ lattice QCD \cite{Chiu:2005zc, Chiu:2007km} 
and $ N_f = 2 $ lattice QCD \cite{Chen:2014hva} respectively.

In this paper, we examine to what extent this scenario holds for $(2+1+1)$-flavors QCD. 
We perform hybrid Monte Carlo simulation of lattice QCD with $ N_f = 2+1+1 $ optimal domain-wall quarks 
\cite{Chiu:2002ir,Chiu:2015sea} on a $ 32^3 \times 64 $ lattice with lattice spacing $ a \simeq 0.063 $~fm, 
keeping $ \s $ and $ \c$ quarks at their physical masses, while $ \u/\d $ sea quarks are unphysically heavy, 
with the unitary pion mass $ M_\pi \simeq 280 $~MeV and $ M_\pi L \simeq 3 $. 
We compute the point-to-point quark propagators, and measure the time-correlation functions of 
meson and baryon interpolators. The mass spectra of the lowest-lying mesons and baryons 
containing $ \s $ and $ \c $ quarks are in good agreement with the high energy experimental values,  
together with the predictions of the charmed baryons which have not been observed in experiments.

\section{Hybrid Monte Carlo Simulation of $N_f=2+1+1$}

First, we point out that, for the domain-wall fermion, to simulate $ N_f = 2 +1 + 1 $ amounts to simulate 
$ N_f = 2 + 2 + 1 $, according to the identity
\bea
\label{eq:Nf2p1p1}
&& \left( \frac{\det \Dodwf(m_{u/d})}{\det \Dodwf(m_{PV})} \right)^2  
   \frac{\det \Dodwf(m_s)}{\det \Dodwf(m_{PV})}   
   \frac{\det \Dodwf(m_c)}{\det \Dodwf(m_{PV})}    \\ 
&=& 
\label{eq:Nf2p2p1_A}
\left( \frac{\det \Dodwf(m_{u/d})}{\det \Dodwf(m_{PV})} \right)^2  
\left( \frac{\det \Dodwf(m_s)}{\det \Dodwf(m_{PV})} \right)^2  
\frac{\det \Dodwf(m_c)}{\det \Dodwf(m_{s})}       \\
&=&
\label{eq:Nf2p2p1_B}
\left( \frac{\det \Dodwf(m_{u/d})}{\det \Dodwf(m_{PV})} \right)^2  
\left( \frac{\det \Dodwf(m_c)}{\det \Dodwf(m_{PV})} \right)^2  
\frac{\det \Dodwf(m_s)}{\det \Dodwf(m_{c})},  
\eea
where $ \Dodwf(m_q) $ denotes the domain-wall fermion operator with bare quark mass $ m_q $,  
and $ m_{PV} $ the mass of the Pauli-Villars field. Since the simulation of 2-flavors 
is much faster than the simulation of one-flavor, it is better to simulate $ N_f = 2 + 2 + 1 $   
than $ N_f = 2 + 1 + 1 $.
Whether (\ref{eq:Nf2p2p1_A}) is more efficient than (\ref{eq:Nf2p2p1_B}) or vice versa 
depends on the computational platform, the algorithm, the lattice size, and the parameters of the action.  
In this work, we choose (\ref{eq:Nf2p2p1_B}) for our HMC simulations. 

For the gluon fields, we use the Wilson plaquette gauge action at $ \beta = 6/g_0^2 = 6.20 $.  
For the two-flavor parts, we use the pseudofermion action for 2-flavors lattice QCD 
with the optimal domain-wall quarks, as defined in Eq. (14) of Ref. \cite{Chiu:2013aaa}.
For the one-flavor part, we use the exact pseudofermion action for one-flavor domain-wall fermion, 
as defined by Eq. (23) of Ref. \cite{Chen:2014hyy}. 
The parameters of the pseudofermion actions are fixed as follows. 
For the domain-wall fermion operator $\Dodwf(m_q) $ defined in Eq. (2) of Ref. \cite{Chiu:2013aaa}, 
we fix $ c = 1, d = 0 $ (i.e., $ H = H_w $), $ m_0 = 1.3 $, $ N_s = 16 $, 
and $ \lambda_{max}/\lambda_{min} = 6.20/0.05 $.  
Note that the optimal weights $ \{ \omega_s, s=1,\cdots,N_s \} $ are different for the 2-flavors action 
and the one-flavor action. For the 2-flavors action, $ \omega_s $ are computed
according to Eq. (12) in Ref. \cite{Chiu:2002ir} such that 
the effective 4D Dirac operator is exactly equal to the Zolotarev optimal rational approximation of 
the overlap Dirac operator with bare quark mass $ m_q $.
For the one-flavor action, $ \omega_s $ are computed according to Eq. (9) in Ref. \cite{Chiu:2015sea}, 
which are the optimal weights satisfying the $R_5 $ symmetry, giving the approximate sign function  $ S(H) $
of the effective 4D Dirac operator satisfying the bound $ 0 < 1-S(\lambda) \le 2 d_Z $ 
for $ \lambda^2 \in [\lambda_{min}^2, \lambda_{max}^2] $,  
where $ d_Z $ is the maximum deviation $ | 1- \sqrt{x} R_Z(x) |_{\rm max} $ of the 
Zolotarev optimal rational polynomial $ R_Z(x) $ of $ 1/\sqrt{x} $ 
for $ x \in [1, \lambda_{max}^2/\lambda_{min}^2] $.    

We perform the HMC simulation of (2+1+1)-flavors QCD on the $ L^3 \times T = 32^3 \times 64$ lattice, 
with the $\u/\d$ quark mass $ m_{u/d} a = 0.005$, 
the strange quark mass $ m_s a = 0.04 $, and the charm quark mass $ m_c a = 0.55 $, 
where the masses of $\s$ and $\c$ quarks are fixed by the masses of the vector mesons
$ \phi(1020) $ and $ J/\psi(3097) $ respectively. 
The algorithm for simulating 2-flavors of optimal domain-wall quarks has been outlined in Ref. \cite{Chiu:2013aaa},
while the exact one-flavor algorithm (EOFA) for domain-wall fermions has been presented in Ref. \cite{Chen:2014hyy}.
Here we note that EOFA outperforms the rational hybrid Monte Carlo algorithm (RHMC) \cite{Clark:2006fx}, 
no matter in terms of the memory consumption or the speed \cite{Chen:2014bbc,Murphy:2016ywx}.

In the molecular dynamics, we use the Omelyan integrator \cite{Omelyan:2001}, 
and the Sexton-Weingarten multiple-time scale method \cite{Sexton:1992nu}.
Moreover, we introduce an auxiliary heavy fermion field with mass $ m_H $ ($ m_q \ll m_H \ll m_{PV} $)
similar to the case of the Wilson fermion \cite{Hasenbusch:2001ne}, the so-called mass preconditioning. 
For the 2-flavors parts, mass preconditioning is only applied to 
the $\u/\d$ quark factor of (\ref{eq:Nf2p2p1_B}) with $ m_H a = 0.1 $.
For the one-flavor part, a novel mass preconditioning has been devised for the EOFA \cite{Chen:2017aaa}, 
which is $\sim 20\% $ faster than the mass preconditioning we have used in 
Refs. \cite{Chen:2014hyy, Chen:2014bbc}. Also, based on the fact that in EOFA 
the fermion force of the $ \phi_1 $ field is much smaller than that of the $ \phi_2 $ field, 
the gauge momentum updating by these two forces can be set at two different time scales. 
Furthermore, we have developed a generalized multiple-time scale method 
with the flexibility of assigning an arbitrary updating time interval to any fermion force  
provided that the updating time interval of the gauge force 
is an integer multiple of the lowest common multiplier (LCM) of the updating intervals of all fermion forces. 
This feature is essential for tuning the parameters to attain optimal efficiency. 
The details of our simulations will be presented in a forthcoming long paper \cite{Chiu:2017bbb}.

We generate the initial 460 trajectories with two Nvidia GTX-TITAN cards 
(each with device memory $ \ge 6 $~GB). 
After discarding the initial 300 trajectories for thermalization, we sample one configuration
every 5 trajectories, resulting 32 ``seed" configurations. 
Then we use these seed configurations as the initial configurations for 32 independent simulations on 32 nodes, 
each of two Nvidia GTX-TITAN cards.  
Each node generates $~50-85$ trajectories independently, and   
all 32 nodes accumulate a total of 2000 trajectories. 
From the saturation of the binning error of the plaquette, as well as
the evolution of the topological charge, 
we estimate the autocorrelation time to be around 5 trajectories. 
Thus we sample one configuration every 5 trajectories, 
and obtain a total of $400$ configurations for physical measurements.

\begin{figure*}[tb]
\begin{center}
\begin{tabular}{@{}c@{}c@{}}
\includegraphics*[height=6.2cm,clip=true]{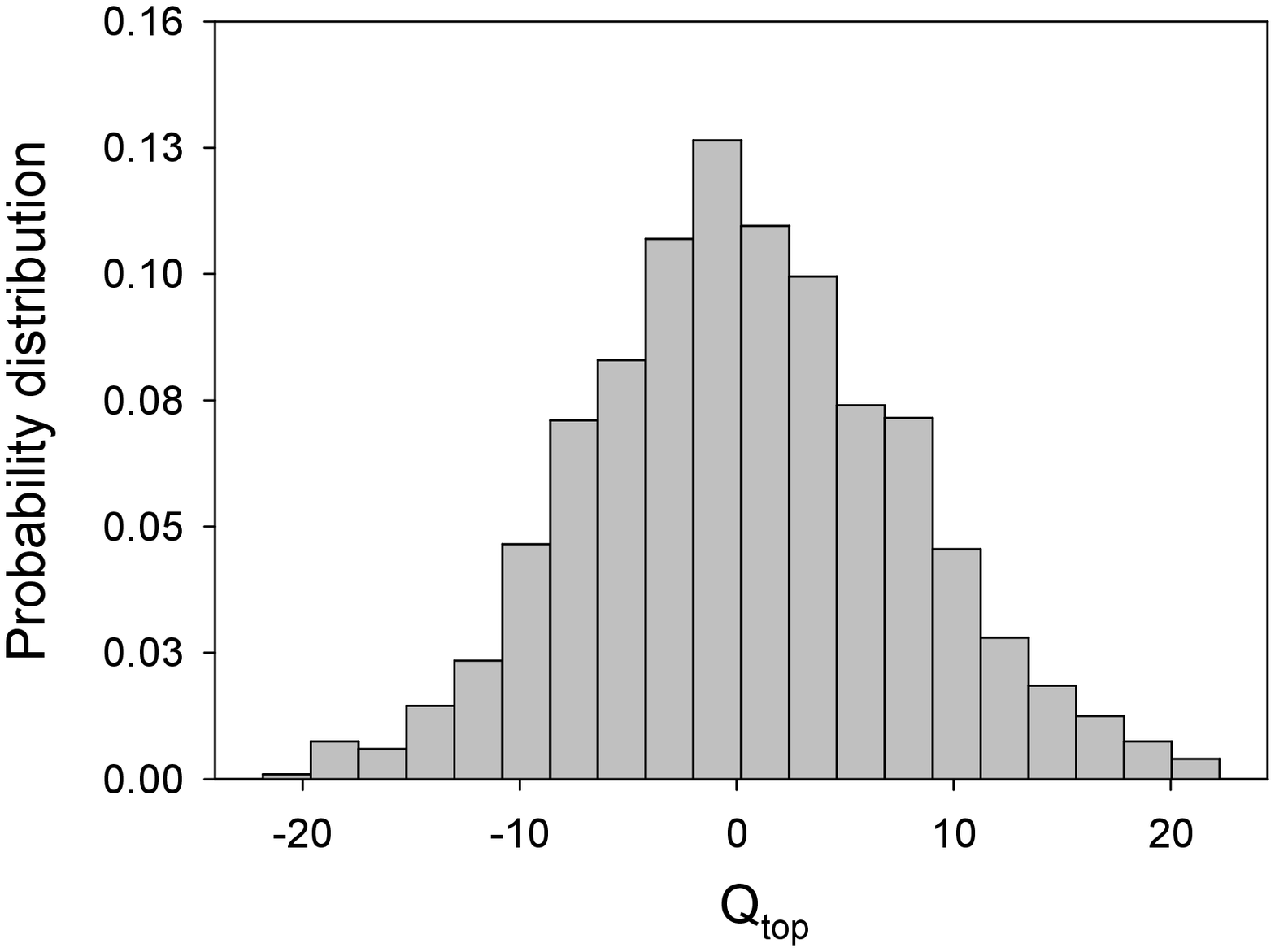}
&
\includegraphics*[height=6.0cm,clip=true]{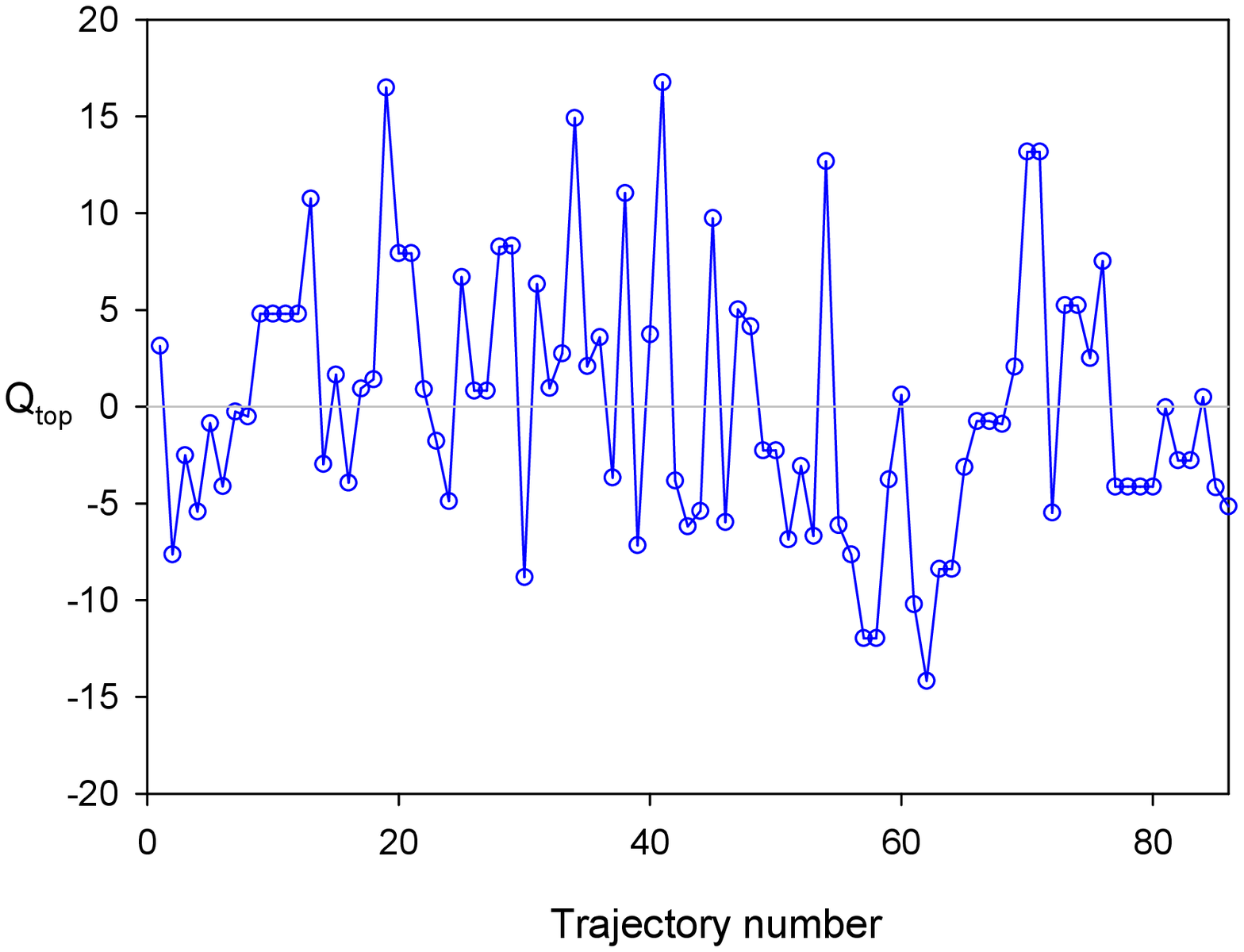}
\\ (a) & (b)
\end{tabular}
\caption{(a) Histogram of topological charge distribution of 2000 trajectories.
(b) Evolution of topological charge in a single stream, one of the 32 streams of independent HMC simulations.  
}
\label{fig:Q_hist}
\end{center}
\end{figure*}

In Fig.~\ref{fig:Q_hist}, we plot the histogram of the topological charge ($ Q_{top} $)   
distribution of 2000 trajectories, together with the evolution of  
the topological charge in one of the 32 streams of independent HMC simulations.  
Evidently, the probability distribution of $ Q_{top} $ behaves like a Gaussian, and 
the HMC simulation in each stream samples all topological sectors ergodically. 
Here the topological charge is measured according to the formula 
$ Q_{top} = \sum_x \epsilon_{\mu\nu\lambda\sigma} \tr[ F_{\mu\nu}(x) F_{\lambda\sigma}(x) ]/(32 \pi^2) $, 
where the matrix-valued field tensor $ F_{\mu\nu}(x) $ is obtained from the four plaquettes  
surrounding $ x $ on the ($\hat\mu,\hat\nu$) plane. Even though this topological charge 
is not exactly equal to an integer, it gives a qualitative picture to demonstrate that 
our HMC simulation samples all topological sectors ergodically. 
For a rigorous determination of the topological charge and susceptibility, 
it requires to project the zero modes of the overlap Dirac operator \cite{Neuberger:1997fp,Narayanan:1994gw}
for each gauge configuration, 
which is beyond the scope of this paper.     

To determine the lattice scale, we use the Wilson flow \cite{Narayanan:2006rf,Luscher:2010iy} with the condition
\bea
\label{eq:t0}
\left. \{ t^2 \langle E(t) \rangle \} \right|_{t=t_0} = 0.3,
\eea
and obtain $ \sqrt{t_0}/a = 2.2737(19) $ for 400 configurations. 
Using $ \sqrt{t_0} =  0.1416(8) $~fm obtained by 
the MILC Collaboration for the $(2+1+1)$-flavors QCD \cite{Bazavov:2015yea}, 
we have $ a^{-1} = 3.167 \pm 0.018 $~GeV.

We compute the valence quark propagator of the 4D effective Dirac operator  
with the point source at the origin, 
and with the mass and other parameters exactly the same as those of the sea quarks. 
First, we solve the following linear system with mixed-precision conjugate gradient algorithm,  
for the even-odd preconditioned ${\cal D} $ (see Eq. (12) in Ref. \cite{Chiu:2013aaa}), 
\bea
\label{eq:DY}
{\cal D}(m_q) |Y \rangle = {\cal D}(m_{PV}) B^{-1} |\mbox{source vector} \rangle, 
\eea
where $ B^{-1}_{x,s;x',s'} = \delta_{x,x'}(P_{-}\delta_{s,s'}+P_{+}\delta_{s+1,s'}) $
with periodic boundary conditions in the fifth dimension.
Then the solution of (\ref{eq:DY}) gives the valence quark propagator  
\BAN
\label{eq:v_quark}
(D_c + m_q)^{-1}_{x,x'} = r \left( 1 - r m_q \right)^{-1} \left[ (BY)_{x,1;x',1} - \delta_{x,x'} \right].   
\EAN
Each column of the quark propagators is computed with 2 Nvidia GTX-TITAN GPUs in one computing node,  
attaining more than one Teraflops/sec (sustained).    

To measure the chiral symmetry breaking due to finite $N_s$, we compute the residual mass
according to the formula \cite{Chen:2012jya}
\BAN
\label{eq:Mres}
m_{res}=\frac{\left< \tr(D_c + m_q)^{-1}_{0,0} \right>_U}{\left< \tr[\gamma_5 (D_c + m_q) \gamma_5 (D_c+m_q)]^{-1}_{0,0} \right>_U}-m_q,
\EAN
where
$ (D_c + m_q)^{-1} $ denotes the valence quark propagator with $ m_q $ equal to the sea-quark mass, 
tr denotes the trace running over the color and Dirac indices, and the brackets $ \left< \cdots \right>_U $
denote the averaging over the gauge ensemble. 
For the 400 gauge configurations generated by HMC simulation of lattice QCD with 
$ N_f = 2 + 1 + 1 $ optimal domain-wall quarks, the residual masses 
of $ \u/\d$, $\s $, and $\c $ quarks are listed in Table \ref{tab:mres}.
We see that the residual mass of the $ \u/\d $ quark is $\sim 1.2$\% of its bare mass, amounting to 
$0.19(4) $~MeV, which is expected to be much smaller than other systematic uncertainties. 
The residual masses of $ \s $ and $ \c $ quarks are even smaller, $ 0.11(3)$~MeV, and $0.07(3) $~MeV 
respectively.  

\begin{table}
\begin{center}
\caption{The residual masses of $ \u/\d $, $\s $, and $ \c $ quarks.}
\vspace{0.5cm}
\begin{tabular}{cccc}
quark & $m_q a $ & $ m_{res} a $ & $ m_{res}$~[MeV]   \\
\hline
\hline
$\u/\d$ & 0.005  & $ (6.0 \pm 1.2) \times 10^{-5} $ & 0.19(4)   \\ 
$\s$    & 0.040  & $ (3.6 \pm 1.1) \times 10^{-5} $ & 0.11(3)   \\ 
$\c$    & 0.550  & $ (2.2 \pm 1.0) \times 10^{-5} $ & 0.07(3)   \\ 
\hline
\end{tabular}
\label{tab:mres}
\end{center}
\end{table}

\section{Mass Spectrum of Hadrons containing $ \s $ and $ \c $ quarks}

One of the main objectives of lattice QCD is to extract the mass spectrum of QCD nonperturbatively from the 
first principles. Even though our $ N_f = 2+1+1 $ gauge ensemble is generated with 
unphysically heavy $\u/\d$ quarks (with $ M_\pi \simeq 280$~MeV), 
we suspect that the mass spectrum and the decay constants of hadrons 
containing $\c$ and $\s$ quarks may turn out to be in good agreement with high energy experimental results, 
as we have observed in $ N_f = 0 $ lattice QCD \cite{Chiu:2005zc, Chiu:2007km} 
and $ N_f = 2 $ lattice QCD \cite{Chen:2014hva} respectively. 
In the following, we examine to what extent this scenario is realized in the spectrum of 
lattice QCD with $N_f = 2+1+1 $ optimal domain-wall quarks. 

Following our previous studies \cite{Chiu:2005zc,Chiu:2007km,Chen:2014hva}, we construct 
quark-antiquark interpolators for mesons, and 3-quarks interpolators for baryons, 
and measure their time-correlation functions using the  
point-to-point quark propagators computed with the same 
parameters ($N_s = 16 $, $m_0 = 1.3 $, $ \lambda_{max}/\lambda_{min} = 6.20/0.05 $) 
and masses ($ m_{u/d} a = 0.005, m_s a = 0.04, m_c a = 0.55 $) of the sea quarks, 
where $m_s$ and $m_c$ are fixed by the masses of the vector mesons
$ \phi(1020) $ and $ J/\psi(3097) $ respectively. 
Then we can extract the mass of the lowest-lying hadron state from the time-correlation function, following the 
procedures outlined in Refs. \cite{Chiu:2005zc,Chiu:2007km,Chen:2014hva}. 


\begin{table}
\begin{center}
\caption{The mass spectrum of the lowest-lying $ \cbar \Gamma \c $
         meson states obtained in this work, in comparison with the PDG values.}
\vspace{0.5cm}
\begin{tabular}{cc|cccc}
$ \Gamma $ & $ J^{PC} $ 
           & $ [t_1,t_2] $ & $\chi^2$/dof & Mass[MeV] & PDG \\
\hline
\hline
$ \Id $ & $ 0^{++} $
                     & [12,19] & 0.69 & 3417(12)(8) & $ \chi_{c0}(3415) $ \\
$ \gamma_5 $ & $ 0^{-+} $
                          & [10,17] & 0.73 & 2980(7)(6) & $ \eta_c(2983) $ \\ 
$ \gamma_i $ & $ 1^{--} $ 
                          & [13,25] & 0.70 & 3097(3)(4) & $ J/\psi(3097) $ \\
$ \gamma_5\gamma_i $ & $ 1^{++} $ 
                                  & [14,21] & 0.90 & 3511(21)(8) & $ \chi_{c1}(3510) $ \\
$ \epsilon_{ijk} \gamma_j \gamma_k $ 
                             & $ 1^{+-} $ 
                             & [11,27] & 0.68 & 3525(13)(5) & $h_c(3525)$ \\
\hline
\end{tabular}
\label{tab:cbar-c}
\end{center}
\end{table}

The mass spectra of the lowest-lying states of the charmonium ($\cbar \c $) and the $ \cbar \s $ mesons 
are summarized in Tables \ref{tab:cbar-c}-\ref{tab:cbar-s}.  
The first column is the Dirac matrix in the meson interpolator $ \Qbar \Gamma \q $. 
The second column is $ J^{PC} $ of the state. 
The third column is the time interval $ [t_1, t_2] $ for  
fitting the data of the time-correlation function $ C_\Gamma(t) $ to the formula   
\bea
\label{eq:single_meson}
\frac{z^2}{2 M a} [ e^{-M a t} + e^{- M a(T-t)} ],   
\eea
to extract the meson mass $ M $ and the amplitude $ z=|\langle H|\Qbar \Gamma \q |0\rangle| $, 
where $ H $ denotes the lowest-lying meson state with zero momentum, 
and the excited states have been neglected.
We use the correlated fit throughout this work. 
The fifth column is the mass $ M $ of the state, where the first 
error is statistical, and the second is systematic error.
Here the statistical error is estimated using the jackknife method 
with the bin-size of which the statistical error saturates, 
while the systematic error is estimated based on all fittings 
satisfying $ \chi^2/\mbox{dof} \le 1.1 $ and $ |t_2 - t_1| \ge 6 $ with
$ t_1 \ge 10 $ and $ t_2 \le 30 $. 
The last column is the corresponding state in high energy experiments,  
with the PDG mass value \cite{Olive:2016xmw}.
Evidently, our results of the mass spectra of the lowest-lying states of the 
charmonium and the $ \cbar s $ mesons are in good agreement with the PDG values. 

For the charmonium, we note that our result of the hyperfine splitting ($M_{J/\Psi} - M_{\eta_c}$) is   
$ 117(8)(7) $ MeV, in good agreement with the PDG value $ 114 $ MeV. 
 
\begin{table}
\begin{center}
\caption{The mass spectrum of the lowest-lying $\cbar\Gamma\s $ meson states
         obtained in this work, in comparison with the PDG values.}
\vspace{0.5cm}
\begin{tabular}{cc|cccc}
$ \Gamma $ & $ J^{P} $ 
           & $ [t_1,t_2] $ & $\chi^2$/dof & Mass[MeV] & PDG \\
\hline
\hline
$ \Id $ & $ 0^{+} $
                     & [17,23] & 0.70 & 2317(15)(5) & $ D^*_{s0}(2317) $ \\
$ \gamma_5 $ & $ 0^{-} $
                          & [15,20] & 0.80 & 1967(3)(4) & $ D_s(1968) $  \\
$ \gamma_i $ & $ 1^{-} $ 
                          & [12,24] & 0.15 & 2112(4)(7) & $ D^*_s(2112) $  \\ 
$ \gamma_5\gamma_i $ & $ 1^{+} $ 
                                  & [13,20] & 0.96 & 2463(13)(9) & $ D_{s1}(2460) $  \\
$ \epsilon_{ijk} \gamma_j \gamma_k $ 
                             & $ 1^{+} $ 
                             & [10,18] & 0.62 & 2536(12)(4) & $D_{s1}(2536)$  \\
\hline
\end{tabular}
\label{tab:cbar-s}
\end{center}
\end{table}

For the $ \cbar \s $ meson states in Table \ref{tab:cbar-s}, their masses are in good agreement with the 
experimental values, implying that they are conventional meson states composed of valence quark-antiquark, 
interacting through the gluons with the quantum fluctuations of $ (\u, \d, \s, \c) $ quarks in the sea. 
It is interesting to see that the masses of the scalar meson $ D^{*}_{s_0}(2317) $, 
and the axial-vector mesons $ D_{s1}(2460) $ and $D_{s1}(2536)$ 
can be obtained with quark-antiquark interpolators, 
without invoking 4-quark interpolators like $ D K $ and $ D^*K $. 
We note that a recent study \cite{Bali:2015lka}
of $ N_f = 2+1 $ lattice QCD with nonperturbatively improved Wilson-clover fermions   
and the same fermion action for the valence quarks, using quark-antiquark interpolators, 
also obtained the masses of the lowest-lying $ \cbar \s $ meson states compatible with the experimental values. 
   
Note that in the physical limit, $ D^*_{s_0}(2317) $ is about 41 MeV below the $DK$ threshold, 
and $ D_{s1}(2460) $ is 44 MeV below the $D^*K $ threshold, while  
$ D_{s1}(2536) $ is 32 MeV above the $D^*K $ threshold. 
Thus it seems to be necessary to consider the effects of the nearby scattering states, e.g., by incorporating  
4-quark interpolators like $ D K $ and $ D^*K $. 
However, for our gauge ensemble, the $DK$ threshold is about 156 MeV above the $ \cbar\s $ scalar meson state, 
and the $D^*K$ threshold 
is more than 220 MeV and 146 MeV above the $ \cbar\s $ axial-vector meson states. 
Moreover, since the time-correlation function is well fitted to the form of 
single meson state (\ref{eq:single_meson}) on plateaus with $ |t_1 - t_2 | \ge 6 $, 
this implies that the ratios 
\BAN
&& \frac{|\langle DK |\cbar\s|0\rangle|^2}{|\langle D^*_{s0}({\text{scalar}})|\cbar\s|0\rangle|^2} 
  \cdot e^{-(M_D + M_K - M_{\text{scalar}})t} \ll 1, \\   
&& \frac{|\langle D^*K |\cbar \gamma_5 \gamma_i\s|0\rangle|^2}
     {|\langle D_{s1}({{\text{axial-vector}}})|\cbar \gamma_5 \gamma_i\s|0\rangle|^2} 
  \cdot e^{-(M_{D^*} + M_K - M_{\text{axial-vector}})t} \ll 1,  \\ 
&& \frac{|\langle D^*K |\cbar \epsilon_{ijk} \gamma_j \gamma_k \s|0\rangle|^2}
     {|\langle D'_{s1}({\text{axial-vector}}) |\cbar \epsilon_{ijk} \gamma_j \gamma_k \s|0\rangle|^2}
 \cdot e^{-(M_{D^*}+M_K-M'_{\text{axial-vector}})t} \ll 1,  
\EAN
are much less than one (at least for $ t \in [10,54] $), for our gauge ensemble.
Nevertheless, it is still interesting to check whether the masses of these states would be   
affected by the threshold effects, by incorporating 4-quark interpolators 
$ D K $ and $ D^* K $, and performing variational analysis   
on the correlation matrices of both 2-quark and 4-quark interpolators, 
similar to the study in Ref. \cite{Lang:2014yfa}, especially for the gauge ensembles approaching the 
physical limit.

Next, we turn to the baryons with $ \s $ and $ \c $ quarks,  
$ \Omega $, $ \Omega_c $, $\Omega_{cc} $, and $ \Omega_{ccc} $. 
Following the notations in our previous study \cite{Chiu:2005zc}, 
their interpolating operators are 
$ (\s C \gamma_\mu \s) \s $,   
$ [\c (C \gamma_5) \s] \s $,   
$ (\c C \gamma_\mu \s) \s $,   
$ [\c C \gamma_5 \s] \c $,   
$ (\c C \gamma_\mu \s) \c $,   
and $ (\c C \gamma_\mu \c) \c $.   
The time-correlation function of any baryon interpolator $ B $ is defined as 
$ 
C_{\alpha\beta}(t) = \sum_{\vec{x}} \langle B_{x\alpha} \bar B_{0\beta} \rangle, 
$
which can be expressed in terms of quark propagators. 

For baryon interpolating operator like $ B^\mu = (\q_1 C \gamma_\mu \q_2) \q_3 $, 
spin projection is required to extract the $ J=3/2 $ state, since it also overlaps 
with the $ J=1/2 $ state. The spin $J=3/2$ projection for the time-correlation function reads 
\BAN
C^{3/2}_{ij}(t) &=& \sum_{k=1}^3 \left(\delta_{ik} - \frac{1}{3} \gamma_i \gamma_k \right) C^{kj}(t), 
\EAN
where
$ C^{kj}(t) = \sum_{\vec{x}} \langle B^{k}(\vec{x},t) \overline{B}^j(\vec{0},0) \rangle $.
Then the mass of the $ J=3/2^\pm $ state can be extracted from
any one of the 9 possibilities ($ i,j = 1,2,3 $) of $C_{ij}^{3/2}(t) $. 
To enhance the statistics, we use $\sum_{i=1}^3 C_{ii}^{3/2}(t)/3 $ to extract the mass of the $ J=3/2 $ state. 

Following the procedures outlined in our previous study \cite{Chiu:2005zc}, 
we obtain the masses of $ \Omega $, $ \Omega_c $, $ \Omega_{cc} $ and $ \Omega_{ccc} $, as summarized 
in Table \ref{tab:omega}. The mass value in the fifth column is obtained by correlated fit, 
where the first error is statistical, and the second is systematic error.
Here the statistical error is estimated using the jackknife method 
with the bin-size of which the statistical error saturates, 
while the systematic error is estimated based on all fittings 
satisfying $ \chi^2/\mbox{dof} \le 1.2 $ and $ |t_2 - t_1| \ge 5 $ with
$ t_1 \ge 10 $ and $ t_2 \le 30 $. 
Evidently, the masses of $ \Omega(3/2^+) $, $ \Omega(3/2^-) $, $ \Omega_c(1/2^+) $, 
and $ \Omega_c(3/2^+) $ are in good agreement with the PDG values in the last column. 
For $ \Omega_c(1/2^-) $, $\Omega_c(3/2^-)$, $ \Omega_{cc}(1/2^{\pm}) $, $ \Omega_{cc} (3/2^{\pm}) $ 
and $ \Omega_{ccc}(3/2^{\pm}) $, they have not been observed in experiments, 
thus their masses in Table \ref{tab:omega} serve as predictions of 
lattice QCD with $ N_f = 2+1+1 $ domain-wall quarks.   

Comparing the spectra of charmed baryons in Table \ref{tab:omega}
with those in our quenched study \cite{Chiu:2005zc}, we see that 
the masses of $ \Omega_c(1/2^{\pm}) $, $ \Omega_c(3/2^{\pm}) $ and $ \Omega_{cc}(3/2^{\pm}) $ 
are in agreement between the cases of $ N_f = 0 $ and $ N_f = 2 + 1 + 1 $,     
while for $ \Omega_{cc}(1/2^{\pm}) $ and $ \Omega_{ccc}(3/2^{\pm}) $, their masses in $ N_f = 2+1+1 $ QCD are 
$\sim 100 $~MeV heavier than their counterparts in the $ N_f = 0 $ QCD. 

It is interesting to point out that the mass of $ \Omega_c(3/2^+) $ was predicted to be 2756(32)~MeV 
in our quenched study \cite{Chiu:2005zc}, before it was observed by the Belle Collaboration in 2009, 
with the measured mass $2765.9 \pm 2.0 $~MeV \cite{Solovieva:2008fw}.
In other words, for lattice QCD with exact chiral symmetry, even in the quenched approximation, 
it can give the mass spectra of heavy hadrons reliably.
This scenario also holds for heavy mesons, e.g.,  
in our quenched study of mesons containing $ \b $, $\c$, and $\s$ quarks \cite{Chiu:2007km},  
we predicted the mass of $\eta_b$ to be $ 9383(4)(2) $~MeV, 
before $\eta_b $ was discovered by the BaBar Collaboration in 2008 \cite{Aubert:2008ba}, 
with the measured mass $ 9388.9^{+3.1}_{-2.3} \pm 2.7 $~MeV.   
 
We note that there are several recent lattice studies of the mass spectra of charmed baryons 
(see, e.g., Refs. \cite{Bali:2015lka,Padmanath:2015jea,Brown:2014ena,Briceno:2012wt,Liu:2009jc}), 
in the framework of $ N_f = 2 $, 2+1, and $ 2+1+1 $ lattice QCD, with  
different fermion actions for the $\c$ quark, and/or the $ \c $ quark is absent in the sea.  
A detailed review of lattice results of charmed baryons is beyond the scope of this paper.

\begin{table}[htb]
\caption{
The mass spectrum of baryon states containing $\s$ and $\c$ quarks obtained in this work.
The last column is from the listings of Particle Data Group \cite{Olive:2016xmw}, 
where $ J^P $ has not been measured for all entries.
}
\begin{tabular}{cc|cccc}
Baryon         & $ J^P $ & $ [t_1,t_2] $ & $\chi^2$/dof & Mass(MeV) & PDG \\
\hline
\hline
$ \Omega $     & $ 3/2^+ $ & [10, 20] & 1.12 & 1680(18)(20)  &  1672  \\
$ \Omega $     & $ 3/2^- $ & [12, 17] & 0.33 & 2248(51)(44)  &  2250  \\
\hline
$ \Omega_{c} $  & $ 1/2^+ $ & [18,30] & 0.74 & 2695(24)(15) &  2695  \\
$ \Omega_{c} $  & $ 1/2^- $ & [14,22] & 0.91 & 3015(29)(34) &        \\
$ \Omega_{c} $  & $ 3/2^+ $ & [18,30] & 1.13 & 2781(12)(22) &  2766  \\
$ \Omega_{c} $  & $ 3/2^- $ & [14,21] & 1.10 & 3210(35)(31) &        \\
\hline
$ \Omega_{cc} $    & $ 1/2^+ $ & [25,30] & 0.90 & 3712(25)(32) &        \\
$ \Omega_{cc} $    & $ 1/2^- $ & [14,20] & 1.06 & 4148(9)(34) &        \\
$ \Omega_{cc} $    & $ 3/2^+ $ & [25,30] & 0.34 & 3785(28)(36) &        \\
$ \Omega_{cc} $    & $ 3/2^- $ & [14,20] & 0.93 & 4200(32)(26) &        \\
\hline
$ \Omega_{ccc} $   & $ 3/2^+ $ & [23,28] & 0.90 & 4766(5)(11) &     \\
$ \Omega_{ccc} $   & $ 3/2^- $ & [17,26] & 1.12 & 5168(37)(51) &     \\
\hline
\end{tabular}
\label{tab:omega}
\end{table}

\section{Summary and Concluding Remarks}

In this paper, we present the first study of lattice QCD with $N_f=2+1+1$ domain-wall quarks.
Using 64 Nvidia GTX-TITAN GPUs evenly distributed on 32 nodes, 
we perform the HMC simulation on the $ 32^3 \times 64 \times 16 $ lattice, with lattice spacing $ a \sim 0.063 $~fm. 
Even though the mass of $ \u/\d $ quarks is unphysically heavy (with unitary pion mass $\sim 280 $~MeV), 
the masses of hadrons containing $ \c $ and $ \s $ quarks turn out in good agreement with the experimental values, 
as summarized in Tables \ref{tab:cbar-c}-\ref{tab:omega}. 
However, extrapolation to the physical limit (with $ M_\pi = 140 $~MeV) is still required, though 
we do not expect significant changes in the mass spectra of hadrons containing $ \s $ and $ \c $ quarks.
Since we have generated only one gauge ensemble, it is impossible for us to perform extrapolation to the 
physical limit, not to mention taking the continuum limit and the infinite volume limit. 
Nevertheless, comparing the mass spectra in Tables \ref{tab:cbar-c}-\ref{tab:omega} to 
those of $N_f=2$ lattice QCD on a $ 24^3 \times 48 $ lattice with $ a \sim 0.063 $~fm \cite{Chang:2017ccc}, 
we conclude that the finite volume uncertainty is much less than the estimated statistical and systematic errors.    
About the discretization error, since the lattice spacing ($a \sim 0.063$~fm) is sufficiently fine, 
and our lattice action is free of $ O(a) $ lattice artifacts, we expect that the discretization error
is also much less than our estimated statistical and systematic errors.     

For the $ \cbar \s $ meson states in Table \ref{tab:cbar-s}, our results  
show that they are conventional meson states composed of valence quark-antiquark, 
interacting through the gluons with the quantum fluctuations of $ (\u,\d,\s,\c) $ quarks in the sea,   
even for the scalar meson $ D^{*}_{s_0}(2317) $, and the axial-vector mesons $ D_{s1}(2460) $ and  
$D_{s1}(2536)$. 

For the mass spectra of baryons in Table \ref{tab:omega}, they are in good agreement with 
the high energy experimental values, together with the predictions 
of the charmed baryons which have not been discovered in experiments.   

To address the challenge of finding out whether there is 
any new physics beyond the SM, it requires to pin down the theoretical uncertainties 
largely from the sector of the strong interaction, 
before one can identify any discrepancies between the experimental 
results and the theoretical values derived from the SM. 
To this end, the latter have to be obtained in a framework 
which preserves all essential features of QCD, i.e., lattice QCD with exact chiral symmetry, 
and also in the unitary limit (with the valence and the sea quarks having the same masses and  
the same Dirac fermion action). 
Otherwise, it is difficult to determine whether any discrepancy between the experimental result
and the theoretical value is due to new physics, or just the approximations 
(e.g., HQET, NRQCD, partially quenched approximation, etc.) one has used. 

To conclude, this work asserts that it is feasible to perform large-scale lattice QCD simulations 
with $ N_f = 2+1+1 $ domain-wall quarks, with good chiral symmetry, and sampling 
all topological sectors ergodically. It provides the ground work for future 
large-scale lattice QCD simulations with dynamical $ (\u,\d,\s,\c,\b) $ domain-wall quarks.

\begin{acknowledgments}
  This work is supported by the Ministry of Science and Technology  
  (Nos.~NSC105-2112-M-002-016, NSC102-2112-M-002-019-MY3), Center for Quantum Science and Engineering 
  (Nos.~NTU-ERP-103R891404, NTU-ERP-104R891404, NTU-ERP-105R891404),  
	and National Center for High-Performance Computing (No. NCHC-j11twc00).  
\end{acknowledgments}

\end{document}